\documentclass[aps,twocolumn,showpacs,amsmath,amssymb,nofootinbib]{revtex4} 

\usepackage{amsmath}
\usepackage[final]{epsfig} 
\usepackage{color} 
\usepackage{graphicx}
\usepackage{dcolumn}
\usepackage{bm}

\begin{document} 
 
\title{Multiplicities and $p_T$ spectra in ultrarelativistic heavy ion collisions from 
a next-to-leading order improved perturbative QCD + saturation + hydrodynamics model} 

\author{R. Paatelainen$^{a,b}$} 
\email{risto.s.paatelainen@jyu.fi}
\author{K.J. Eskola$^{a,b}$}
\email{kari.eskola@phys.jyu.fi}
 \author{H. Holopainen$^{c}$} 
\email{holopainen@fias.uni-frankfurt.de}
\author{K. Tuominen$^{a,b}$}
\email{kimmo.i.tuominen@jyu.fi}
 
\affiliation{
$^a$Department of Physics, P.O.Box 35, FIN-40014 University of Jyv\"askyl\"a, Finland\\ 
$^b$Helsinki Institute of Physics, P.O.Box 64, FIN-00014 University of Helsinki, Finland\\
$^c$Frankfurt Institute for Advanced Studies, Ruth-Moufang-Str. 1, D-60438 Frankfurt am Main, Germany} 
 
\date{\today} 
 
\begin{abstract}
We bring the EKRT framework, which combines perturbative QCD (pQCD) minijet production with gluon saturation and hydrodynamics, to next-to-leading order (NLO) in pQCD as rigorously as possible. We chart the model uncertainties, and study the viability and predictive power of the model in the light of the RHIC and LHC measurements in central $A$+$A$ collisions. 
In particular, we introduce a new set of measurement functions to define the infrared- and collinear-safe minijet transverse energy, $E_T$, in terms of which we formulate the saturation.
We update the framework with the EPS09 NLO nuclear parton distributions (nPDFs), and study the propagation of the nPDF uncertainties into the computed $E_T$, saturation scales and the final-state multiplicities. The key parameters, which need to be fixed using the measurements, are identified, and their correlation is discussed.
We convert the saturated minijet $E_T$ into QCD-matter initial conditions for longitudinally boost-invariant ideal hydrodynamics. We compute the charged-particle multiplicities and identified bulk hadron $p_T$ spectra in 5\% most central Au+Au collisions at RHIC and Pb+Pb at the LHC. We obtain an encouragingly good agreement with the experimental data, simultaneously at RHIC and LHC, showing that the approach has a definite predictive power.
\end{abstract} 
 

\maketitle 

\section{Introduction}
\label{intro}

Different models based on the phenomenon of gluon saturation have offered appealing frameworks to predict and explain the observed charged-particle multiplicities in Pb+Pb collisions at $\sqrt{s_{NN}}=2.76$~GeV at the Large Hadron Collider (LHC), as well as their systematics from the Relativistic Heavy Ion Collider (RHIC) to LHC: for a collection of the LHC predictions, see Ref.~\cite{Abreu:2007kv}, a compilation in Ref.~\cite{Armesto:2008fj}, and a comparison with the first ALICE data in Ref.~\cite{Aamodt:2010pb}. Among these is also the so-called ''EKRT'' model introduced in Ref.~\cite{EKRT}, which combines perturbative QCD minijet production \cite{EKL} with a saturation of gluons\footnote{The concept of saturation was introduced first in \cite{Gribov:1984tu,Mueller:1985wy}, and later in the contexts of the classical gluon fields \cite{McLerran:1993ni} and minijet production \cite{Eskola:1996ce}.} to compute (instead of fitting) the initial conditions for  the hydrodynamical evolution of the system. This model is often referred to as  the ''final-state saturation'' model since 
the gluon saturation here refers to the saturation in gluon production rather than to that in the initial state (color-glass condensate) wave functions of the colliding nuclei. The purpose of this paper is, in addition to the implementation of the next-to-leading-order (NLO) pQCD updates, to improve the conceptual and technical NLO framework of the EKRT modeling, to make an effort to quantify the underlying uncertainties, and to study the predictive power of the approach. We show that after the improvements this fairly robust modeling leads to a very reasonable phenomenology for the particle multiplicities and $p_T$ spectra simultaneously at RHIC and the LHC. 

\subsection{Original pQCD + saturation + hydrodynamics framework}

Let us first look back at the key elements as well as the key predictions in the original EKRT model setups \cite{EKRT,ERRT,EHNRR}. 
The gluon saturation in central (${\bf b}={\bf 0}$) $A$+$A$ collisions was implemented through an uncertainty-relation based geometrical saturation criterion, 
\begin{equation}
N_{AA}(p_0,\sqrt{s_{NN}}, \Delta Y=1,{\bf b}={\bf 0})\times \frac{\pi}{p_0^2} = K_{\rm sat}\pi R_A^2,
\label{eq:saturation}
\end{equation}
where $N_{AA}$ is the number of produced minijets (to a good first approximation all gluons)  \cite{EKL} which fall into the mid-rapidity acceptance window $\Delta Y=1$ with a few-GeV transverse momenta, $p_T\ge p_0\gg \Lambda_{\rm QCD}$. The factors $\pi/p_0^2$ and $\pi R_A^2$ (with $R_A$ the nuclear radius) account for the transverse area occupied by the production of each minijet and that available in a central $A$+$A$ collision, and  the proportionality constant $K_{\rm sat}$ was set to unity. 
In terms of collinear factorization, i.e. DGLAP-evolved nuclear parton distribution functions (nPDFs) $f_{i/A}$, and perturbatively computable leading-order subcross-sections ${\rm d}\hat \sigma^{ij\rightarrow kl}$ the minijet number can be computed as\footnote{We adopt the notation of Refs.~\cite{EHNRR,ET2} here.} 
\begin{equation}
N_{AA}(p_0,\sqrt{s_{NN}}, \Delta Y,{\bf 0})= T_{AA}({\bf 0}) \sigma\langle N \rangle_{\Delta Y,p_0},
\label{eq:NAA}
\end{equation}
where $T_{AA}({\bf 0})=\int d^2{\bf s}\, T_A({\bf s})T_A({\bf s})$ is the standard nuclear overlap function, expressed in terms of the nuclear thickness functions $T_A$, and the hard QCD cross section is (for the parton bookkeeping details, see Ref.~\cite{EHNRR})
\begin{equation}
\begin{split}
\sigma\langle N \rangle_{\Delta Y,p_0} & =  K_{\rm QCD} \int dp_T^2 dy_1 dy_2 \, \tilde S_2(N;p_1,p_2)\,\times \\
& \times\sum_{ijkl} x_1 f_{i/A}(x_1,Q^2) x_2f_{j/A}(x_2,Q^2)
\frac{d\hat \sigma}{d\hat t}^{ij\rightarrow kl}.
\end{split}
\label{eq:sigmaN}
\end{equation}
The ''measurement function'' $\tilde S_2$ \cite{ET1,ET2} above contains the imposed $p_T$ cut-off $p_0$ and rapidity acceptance $\Delta Y$,
\begin{equation}
\tilde S_2(N; p_1,p_2) = \theta(p_T\ge p_0)[\theta(y_1\in \Delta Y) + \theta(y_2\in \Delta Y)],
\label{eq:S2N}
\end{equation}
where the labels 1 and 2 refer to the final state partons and $p_T$ to their (identical) transverse momentum. The functions denoted by $\theta$ are the usual step functions.
The factor $K_{\rm QCD}$ in Eq.~\eqref{eq:sigmaN} above accounts for the NLO corrections. In the original EKRT paper \cite{EKRT} a constant $K_{\rm QCD}=2$ was assumed for simplicity (and since no calculation for these existed) but in the later setups \cite{ERRT,EHNRR} a $\sqrt{s_{NN}}$-dependent $K_{\rm QCD}$ was introduced on the basis of NLO computations of minijet $E_T$ production \cite{ET1,ET2}.

Once the saturation momentum $p_0=p_{\rm sat}$ fulfilling the criterion in Eq.~\eqref{eq:saturation} is found, the amount of transverse energy carried by the minijets (whose $p_T\ge p_{\rm sat}$) into the rapidity acceptance $\Delta Y$ can be computed as 
\begin{equation}
E_T^{AA}(p_0,\sqrt{s_{NN}}, \Delta Y,{\bf 0})= T_{AA}({\bf 0}) \sigma\langle E_T \rangle_{\Delta Y,p_0},
\label{eq:ET}
\end{equation}
where $\sigma\langle E_T \rangle_{\Delta Y,p_0}$ is obtained from Eq.~\eqref{eq:sigmaN} by replacing the measurement function $\tilde S_2(N; p_1,p_2)$ with a new one for $E_T$, 
\begin{equation}
\tilde S_2(E_T; p_1,p_2) = \theta(p_T\ge p_0)p_T[\theta(y_1\in \Delta Y) + \theta(y_2\in \Delta Y)].
\label{eq:S2ET}
\end{equation}

The initial conditions for hydrodynamics are then obtained from the computed $E_T(p_0=p_{\rm sat})$ and saturation scale $p_{\rm sat}$ by assuming that the system thermalizes essentially at formation, $\tau_0=1/p_{\rm sat}$. The average initial energy density becomes then (see \cite{EKRT,ERRT,EHNRR}) $\langle \epsilon \rangle = E_T^{AA}/(\pi R_A^2 \tau_0\Delta Y)$. As discussed in \cite{EKRT}, these initial conditions fix the rapidity density of entropy $dS/d\eta$ in the system, which in turn, with ideal hydrodynamics, is directly proportional to the final-state multiplicity. In the original EKRT framework \cite{EKRT}, where hydrodynamics with only 1-dimensional Bjorken scaling flow was considered, the main prediction was the obtained scaling law of the particle multiplicity, $dN/dy\propto A^{0.92} s_{NN}^{0.19\dots0.20}$, where the binary collision scaling $A^{4/3}$ of independent minijet production was thus tamed to be close to a wounded-nucleon scaling, and where a definite power-law behaviour in $\sqrt{s_{NN}}$ was predicted. Given the robusteness of the model and also that these predictions were prepared before any RHIC data were available, the obtained $\sqrt{s_{NN}}$ scaling turned out to be surprisingly close to the one obtained with the first LHC heavy-ion ALICE data \cite{Aamodt:2010pb}. Another interesting observation in the original EKRT model \cite{EKRT} was that the ratio $E_T^{AA}(p_{\rm sat})/N_{AA}(p_{\rm sat})$ is very close to the ratio $\epsilon(T)/n(T)$ of an ideal massless boson gas, which suggests that further gluon multiplication is not necessary for (kinetic) thermalization. This observation, together with the fact that the system becomes overdense with gluons at saturation, lends support to the early initialization time $1/p_{\rm sat}$ of the hydrodynamical evolution.

In the later, more detailed EKRT setups \cite{ERRT,ENRR,EHNRR}, where the pQCD calculation incluced 
the $\sqrt{s_{NN}}$-dependent $K_{\rm QCD}$ factors \cite{ET1,ET2} mentioned above, the saturated minijet initial conditions served as input for ideal 1+1 D hydrodynamics (central collisions, azimuthal symmetry, longitudinal scaling flow but dynamically evolving transverse flow) once a binary-collision (BC) transverse profile for the energy density,
\begin{equation}
\label{eq:enerden}
\epsilon(s) = T_A(s) T_A(s) \frac{\sigma\langle E_T \rangle_{\Delta Y,p_{\rm sat}}}{\tau_0\Delta Y},
\end{equation}
was assumed. Also a more realistic equation of state, resonance decays and centrality selection were considered in the hydrodynamical description. With such a setup, the charged-particle multiplicities measured in central Au+Au collisions at RHIC at $\sqrt{s_{NN}}= 56,$~130 and 200 GeV were all predicted quite nicely, see the comparison with the data in Refs.~\cite{ERRT,Ruuskanen:2001gs}. Remarkably, given that the prediction in Ref.~\cite{ERRT} was made a decade before the LHC data and before the RHIC $\sqrt{s_{NN}}=200$~GeV data, also the corresponding multiplicity in central $\sqrt{s_{NN}}=2.76$~TeV Pb+Pb collisions at the LHC was  predicted reasonably well.\footnote{{\protect The result obtained by interpolation} from the predictions shown in Ref.~\cite{ERRT} agreed with the ALICE multiplicity (upper limit of the experimental error bar) \cite{Aamodt:2010pb} within 7\%.} 

As discussed in Refs.~\cite{ENRR,EHNRR}, a good agreement with the measured pion and kaon transverse momentum spectra in central Au+Au collisions at RHIC ($\sqrt{s_{NN}}=130 $~and 200 GeV) was also obtained.  
The first EKRT-based results obtained in Ref.~\cite{Systematics}  for charged-hadron $p_T$ spectra in Pb+Pb collisions at the LHC suggest that the framework can be expected to work well also there. 
What should be emphasized with these RHIC and LHC results is the robustness of the framework: due to the pressure, the $PdV$ work, the rapidity density of energy degrades by almost a factor three during the hydrodynamical evolution, so that it is quite nontrivial that a good agreement with the measured $p_T$ spectra follows.

Also the centrality dependence of multiplicities in the EKRT approach has been studied, using the optical Glauber model for the collision geometry and localizing the saturation criterion of Eq.~\eqref{eq:saturation} in the transverse-coordinate plane \cite{EKT1}. As pointed out in Ref.~\cite{Miller:2007ri}, a good agreement with the RHIC data for multiplicity-per-partipant-pair vs.~the number of participants is found when the same (optical) Glauber model is applied also in the data analysis -- for comparison, see Fig.~23(a) in Ref.~\cite{Abelev:2008ab} and Fig.~4 in Ref.~\cite{EKT1}. Elliptic flow predictions in the EKRT framework can be found in Ref.~\cite{KHHET} for RHIC  and  in Refs.~\cite{Niemi:2008ta,Abreu:2007kv} for the LHC.

\subsection{Open questions with the original EKRT setup}

Inspite of the working phenomenology achieved, there are a number of open questions and shortcomings with the original EKRT model setup which we will consider in the present study: 

\textit{(i)} The formulation of the model with saturation of the number of produced gluons, $N_{AA}$ in Eq.~\eqref{eq:saturation}, is problematic, since  the number of gluons cannot be defined in a manner which would be collinear (CL) and infra-red (IR) singularity safe also in NLO pQCD: without introducing an extra resolution scale, how should one count soft gluons and two collinear gluons? Consequently, the factors $K_{\rm QCD}$ in Eq.~\eqref{eq:sigmaN} cannot be computed directly for $N_{AA}$ but they have been extracted from an NLO calculation of transverse energy $E_T^{AA}$ which is a CL/IR-safe quantity and can thus be rigorously defined and computed as in Refs.~\cite{ET1,ET2}. These problems would obviously be avoided if the saturation would be required for $E_T^{AA}$ instead of $N_{AA}$.

\textit{(ii)} There are uncertainties related to the geometrical final-state saturation criterion: It has not been clear whether an explicit strong coupling constant $\alpha_s$ should appear in Eq.~\eqref{eq:saturation} if it describes a fusion of the produced gluons. Also, it has been unclear what is the role of the possible further proportionality constant $K_{\rm sat}$ in Eq.~\eqref{eq:saturation}.
Also, if the softer gluons at $p_T\le p_{\rm sat}$ are produced but they fuse in the overdensely populated phase space, one may ask whether the total transverse energy should still increase with the produced soft gluons? 

\textit{(iii)} As explained above, the hydrodynamical evolution in this framework is initiated with the produced minijet transverse energy and not with their number, computed at $p_0=p_{\rm sat}$ once $p_{\rm sat}$ has been determined on the basis of the minijet number. The whole procedure of computing the produced initial energy densities for hydrodynamics would obviously become much more straightforward if the saturation criterion could be formulated for the produced $E_T^{AA}$ instead of $N_{AA}$.

\textit{(iv)} The saturation criterion in Eq.~\eqref{eq:saturation} is extensive in $\Delta Y$ on the left-hand side but not on the right-hand side (see also the discussion in Ref.~\cite{EKRT2}). Since the previous studies have been made for $\Delta Y=1$, this has not been a problem in practice but obviously also this question calls for an improved formulation of the saturation criterion, where $\Delta Y$ would explicitly appear also on the right-hand side of Eq.~\eqref{eq:saturation}.

\textit{(v)} The earlier EKRT studies (as well as the NLO $E_T$ studies \cite{ET2}) have so far applied only  the EKS98 \cite{EKS98} LO nPDFs in the computation of the minijet cross sections. An NLO update regarding the nPDFs should be done, and also a study of the propagation of the nPDF uncertainties into the hydrodynamic initial conditions and final state multiplicities should be performed.

In this paper, we address the above issues as follows: 
First, we reformulate the gluon saturation in terms of transverse energy per rapidity unit. This solves the first item above by removing the need for the phenomenological factors $K_{\rm QCD}$ for $N_{AA}$, since we can now directly compute the saturating quantity in NLO pQCD. Also the third item is then solved. To address the second item, we show how the geometric saturation criterion without an explicit $\alpha_s$ (or its powers) arises with a more detailed description of the EKRT saturation mechanism. Also the fourth item gets conveniently solved with this reformulation. To address the fifth issue, we bring the NLO computation for minijet $E_T$ production up-to-date by using the EPS09 NLO nPDFs \cite{EPS09}. We also study the nPDF-originating uncertainties in the final-state multiplicities using the EPS09 error sets.

Our aim here is to address also the uncertainties and phenomenological parameters of the approach more concretely than before. In the rigorously computable pQCD part, we discuss the freedom in defining the measurement function(s) for the NLO $E_T$ calculation. In the more phenomenological saturation part, we quantify the uncertainty due to the unknown proportionality constant in the saturation criterion. We will also show how these uncertainties (or rather, freedoms)  are correlated when studying the final-state multiplicities at the LHC and RHIC. At the same time, we chart the predictive power of the improved EKRT framework.  

Once getting the pQCD+saturation part and multiplicities under improved control, we also consider the bulk hadron $p_T$ spectra in central Au+Au collisions at RHIC and Pb+Pb at the LHC. To get an updated EKRT baseline for further improvements, we apply ideal hydrodynamics with a state-of-the art equation of state. It should be emphasized that our goal here is not to tune the model to fit the  $p_T$ spectra at RHIC and LHC as perfectly as possible but, rather, to study whether a reasonable agreement especially with the LHC $p_T$ spectra \cite{Preghenella:2011jv} can be found, by keeping exactly the same parameter setup for RHIC and LHC. We also discuss how the improved understanding of the EKRT framework presented in this paper leaves room also for further improvements such as adding dissipation (viscosity) into the hydrodynamical description.

The rest of the paper is organized as follows:
In Sec.~\ref{sec:updatedEKRT} we present the improvements and updates of the NLO pQCD + saturation framework and specify our hydrodynamical setup. Section~\ref{sec:results} contains the obtained results for the systematics of multiplicities and $p_T$ spectra at RHIC and the LHC. In Sec.~\ref{sec:outlook} we summarize and discuss the further improvements left for future work.

\section{Improved EKRT framework}
\label{sec:updatedEKRT}

\subsection{Saturation in $E_T$}
We take the following new angle in interpreting the saturation in the EKRT framework: instead of a saturation of the number of produced final state gluons we suggest the saturation to take place in transverse energy production when the $3\rightarrow2$ and higher-order $(n\ge 4)\rightarrow 2$ processes start to dominate over the conventional $2\rightarrow2$ processes. The saturation should then not be considered
just as final (or initial) state saturation but all $(n\ge 3)\rightarrow 2$ processes which decrease the gluon number relative to the independent $2\rightarrow 2$ processes, are in effect. Also, as all the higher order processes then effectively reduce the number of final-state gluons, and since the initial state gluons do not carry transverse momentum, the production of $E_T$ from the collisions at $p_T< p_{\rm sat}$ is conjectured to be negligible. 

We thus require that at saturation (which is assumed transversally non-local here), the rapidity densities of the transverse energy
fulfill the condition 
\begin{equation}
\frac{dE_T}{dy}(2\rightarrow 2) \sim \frac{dE_T}{dy}(3\rightarrow 2).
\label{eq:32satET}
\end{equation} 
Considering only the leading-order order cases in $\alpha_s$, we assign a factor $T_A \cdot g$ for each of the incoming gluons ($g$ for gluon PDFs, and $T_A\sim A/(\pi R_A^2)$), the factor $\pi R_A^2$ for the transverse integration $d^2s$ in $T_{AA}$, the appropriate powers of $\alpha_s$, the $p_T$ cut-off scale $p_0$ for the $E_T$, the scale $p_0^{-2}$ for $\sigma(2\rightarrow2)$, and the scale $p_0^{-2}$  to compensate for the fm$^{-2}$ dimension of the extra $T_A$ in the $3\rightarrow 2$ case, we arrive at a scaling law
\begin{equation}
\pi R_A^2(T_Ag)^2\frac{\alpha_s^2}{p_0^2}p_0 \sim \pi R_A^2 (T_Ag)^3\frac{1}{p_0^2}\frac{\alpha_s^3}{p_0^2}p_0.
\label{eq:newsat}
\end{equation} 
At saturation, $p_0=p_{\rm sat}$, this leads to a scaling
\begin{equation}
T_Ag \sim \frac{p_0^2}{\alpha_s}
\label{eq:gsat}
\end{equation} 
for the gluon density probed at saturation.\footnote{Interestingly, a similar relation is traditionally obtained in the CGC framework, see e.g.~Ref. \cite{Mueller:1999wm}.}
Feeding this scaling back to Eq.~\eqref{eq:32satET} gives
\begin{equation}
\frac{dE_T}{dy}(2\rightarrow 2) \propto  R_A^2p_0^3 
\label{eq:ETsat}
\end{equation} 
at saturation.
Thus, we arrive at the following geometrical-like saturation criterion for the average minijet transverse energy produced in a central $A$+$A$ collision:
\begin{equation}
E_T^{AA}(p_0,\sqrt{s_{NN}}, \Delta Y,{\bf 0}) = K_{\rm sat} R_A^2 p_0^3 \Delta Y,
\label{eq:newEKRTsat}   
\end{equation}
where no explicit $\alpha_s$ appears, the rapidity interval $\Delta Y$ (from $dy$) appears also on the right-hand side,
and where the proportionality constant $K_{\rm sat}$ is to be determined on the basis of the measured data at one chosen cms-energy.
This is the saturation conjecture we now test in the following.

\subsection{NLO computation of minijet $E_T$}

The exact NLO formulation of minijet $E_T$ production, introduced originally in Refs.~\cite{ET1,ET2}, is based on collinear factorization and the subtraction method \cite{Kunszt}. As shown in Eq.~\eqref{eq:ET}, the evaluation of the produced $E_T$ in central $A$+$A$ collisions is carried out by computing the first $E_T$ moment $\sigma\langle E_T\rangle_{\Delta Y, p_0}$  of the perturbative $E_T$ distribution of minijets in a single $N$+$N$ collision and including the nuclear collision geometry through the standard nuclear overlap function $T_{AA}({\bf b})$. 
The NLO pQCD framework is described in detail in Ref. \cite{ET2} but we briefly recapitulate the formulation of $\sigma\langle E_T\rangle$ here, to discuss a generalization of the measurement functions for $E_T$.

The semi-inclusive $E_T$ distribution of minijets in a rapidity interval $\Delta Y$ in $N$+$N$ collisions \cite{EKL,ET1,ET2} can be computed to NLO pQCD as 
\begin{equation}
\label{eq:siET}
\begin{split}
&\frac{d\sigma}{dE_T}\Bigg\vert_{\Delta Y,p_0}  = \frac{d\sigma}{dE_T}\Bigg\vert^{2 \rightarrow 2}_{\Delta Y,p_0}  + \frac{d\sigma}{dE_T}\Bigg\vert^{2 \rightarrow 3}_{\Delta Y,p_0}\\
& = \frac{1}{2!}\int d[PS]_2\frac{d\sigma^{2 \rightarrow 2}}{d[PS]_2}S_2 + \frac{1}{3!}\int d[PS]_3\frac{d\sigma^{2 \rightarrow 3}}{d[PS]_3}S_3,
\end{split}
\end{equation}
where the integrations take place in $4-2\epsilon$ spacetime dimensions, and 
the $2\rightarrow 2$ and $2\rightarrow 3$ differential partonic cross sections are denoted as 
\begin{equation}
\label{eq:IETDIST}
\begin{split}
\frac{d\sigma^{2 \rightarrow 2}}{d[PS]_2} & = \frac{d\sigma^{2 \rightarrow 2}}{dp_{T2}dy_1dy_2d\phi_2}\\
\frac{d\sigma^{2 \rightarrow 3}}{d[PS]_3} & = \frac{d\sigma^{2 \rightarrow 3}}{dy_1dy_2dy_3dp_{T2}dp_{T3}d\phi_2d\phi_3}. 
\end{split}
\end{equation}
For the two-parton final state the appropriate kinematical variables are the rapidities $y_1,y_2$, transverse momentum $p_{T2}$ and azimuth angle $\phi_2$. From transverse momentum conservation, we have $p_{T1} = p_{T2}$ and $\phi_1 = \phi_2 + \pi$ as we do not include any intrinsic transverse momentum.
Similarly, for the three-parton final state, the suitable kinematical variables are $y_1,y_2,y_3,p_{T2},p_{T3},\phi_2$ and $\phi_3$, since the transverse momentum conservation fixes 
${\bf p_{T1}}=-({\bf p_{T2}} + {\bf p_{T3}})$.  The measurement functions $S_2=S_2(p_1,p_2)$ and $S_3=S_3(p_1,p_2,p_3)$, which depend on the four-momenta of the final-state partons, define the physical quantity to be computed. In this case this is the $E_T$ distribution of minijets  which fall into a given rapidity acceptance window $\Delta Y$ and which originate from hard (perturbative) collisions where at least an amount $2p_0$ of transverse momentum is produced. Thus, the measurement functions also define what we mean by a hard process here.

To regularize the IR/CL divergencies present in the partonic NLO cross sections, 
we must consider the ultraviolet(UV)-renormalized squared 
$2\rightarrow2$ and $2\rightarrow 3$  scattering matrix elements of order $\alpha_s^3$, in $4-2\epsilon$ dimensions and $\overline{\text{MS}}$ scheme. The divergent terms show an $\epsilon^{-1}$ and $\epsilon^{-2}$ behavior. The full analytical calculation for these matrix elements was done in Ref.~\cite{Ellis}\footnote{Details of some of these rather complicated calculations will be elucidated in \cite{RistoPhD}.}. The explicit cancellation of the IR/CL divergencies takes place only if the three-parton measurement function, $S_3$, reduces to the two-parton one, $S_2$, in the IR and CL limits \cite{Kunszt}, i.e. when one of the final state partons becomes soft, or collinear with any other parton in the process.     

In a hard scattering of partons to NLO, we may have one, two, three or zero minijets in the rapidity acceptance region $\Delta Y$, which here is the mid-rapidity unit, 
\begin{equation}
\Delta Y: \quad \vert y \vert \leq 0.5,\quad 0 \leq \phi \leq 2\pi. 
\end{equation}
All the partons are assumed massless, thus the transverse energy within $\Delta Y$ is the sum of the absolute values $p_{Ti}$ of the transverse momenta of those partons whose rapidities are in $\Delta Y$:
\begin{equation}
\label{ETDEF}
E_T = \epsilon(y_1)p_{T1} + \epsilon(y_2)p_{T2} + \epsilon(y_3)p_{T3},
\end{equation}
where the step function $\epsilon(y_i)$ is defined as
\begin{equation}
\epsilon(y_i) \equiv \left\{ \begin{array}{rcl}
1 & ~\mbox{if} ~ y_i \in \Delta Y  \\ 
0 & \mbox{otherwise.}  
\end{array}\right.
\end{equation}

In the LO and NLO $2 \rightarrow 2$ cases, where the transverse momenta are equal in magnitude, $p_{T1} = p_{T2}$, the hard scatterings can be defined to be those with large enough transverse momentum, regardless of where the partons go in rapidity:  $p_T \geq p_0 \gg \Lambda_{\rm QCD}$, or equivalently, 
\begin{equation}
p_{T1} + p_{T2} \geq 2p_0,
\label{hard2}
\end{equation}
where $p_0$ is a fixed external parameter which does not depend on $\Delta Y$.  
This readily generalizes to the NLO $2 \rightarrow 3$ processes as 
\begin{equation}
p_{T1} + p_{T2} + p_{T3} \geq 2p_0. 
\label{hard3}
\end{equation}

As discussed in Ref.~\cite{ET2}, a possible further element in defining the measurement functions here is that in the $2\rightarrow3$ case we may still restrict the amount of minimum $E_T$ at $\Delta Y$ in an IR/CL safe way: In the $2\rightarrow2$ case, the non-zero $E_T$ in $\Delta Y$  is always larger than $p_0$, while in the $2\rightarrow3$ case we can have (and have plenty of, see \cite{ET2}) hard processes where two partons fall outside $\Delta Y$ and one soft parton inside. At the IR limit in this special case,  we obviously have no $E_T$ in $\Delta Y$ and the usual $2\rightarrow2$ limit is correctly recovered. The other equally well IR/CL safe extreme case is that we could require the $E_T$ in  $\Delta Y$ be at least $p_0$ as always is in the $2\rightarrow 2$ case. The new feature introduced in the present study is that in fact \textit{any minimum amount of $E_T$ between 0 and $p_0$} constitutes an equally good, IR/CL safe restriction for the $E_T$ in $\Delta Y$ which relaxes back to the $2\rightarrow 2$ case at the IR and CL limits.

By combining the definitions of $E_T$ in $\Delta Y$ in Eqs. \eqref{ETDEF}, \eqref{hard2} and \eqref{hard3} together with the definition of the hard scatterings and the possible restriction of $E_T$ discussed above, the IR/CL safe measurement function $S_2$ can now be written down for the $2\rightarrow 2$ scatterings as
\begin{equation}
\label{eq:mf2to2}
\begin{split}
S_2(p_1,p_2) & =  \theta(p_{T1} + p_{T2} \geq 2p_0)\\
&\delta(E_T-\{\epsilon(y_1)p_{T1} + \epsilon(y_2)p_{T2} \})
\end{split}
\end{equation}
and $S_3$ for the $2\rightarrow 3$ scatterings as 
\begin{equation}
\label{eq:mf2to3}
\begin{split}
S_3&(p_1,p_2,p_3) =  \theta(p_{T1} + p_{T2} + p_{T3} \geq 2p_0)\theta(E_T \geq \beta p_0)\\
&\delta(E_T-\{\epsilon(y_1)p_{T1} + \epsilon(y_2)p_{T2} + \epsilon(y_3)p_{T3} \}),
\end{split}
\end{equation}
where the constant $\beta \in [0,1]$. The measurement function $S_3$ above thus generalizes the formulation of Ref.~\cite{ET2}, where only the special cases $\beta =0$ and $1$ were studied. In the IR and CL limits, we can see that $S_3\rightarrow S_2$ as is required for the cancellation of the corresponding divergencies. Importantly, we notice already here that with the parameter $\beta$ we can control how much $E_T$ is allowed to form within a certain formation time $\propto1/p_0$, which in turn affects the hydrodynamical initial energy densities. Thus, $\beta$ is a parameter of this model, whose value is to be determined from the RHIC and/or LHC data as will be explained in Sec. III.

From Eq. \eqref{eq:siET} we now obtain the first moment of the semi-inclusive $E_T$ distribution as
\begin{equation}
\label{eq:sigmaET}
\begin{split}
\sigma\langle E_T \rangle_{p_0,\Delta Y} &\equiv \int_{0}^{\sqrt{s}} dE_T E_T\frac{d\sigma}{dE_T}\Bigg\vert_{p_0,\Delta Y}\\
& = \sigma\langle E_T \rangle_{p_0,\Delta Y}^{2\rightarrow 2} + \sigma\langle E_T \rangle_{p_0,\Delta Y}^{2\rightarrow 3}
\end{split}
\end{equation} 
where, in
\begin{equation}
\sigma\langle E_T \rangle_{p_0,\Delta Y}^{2\rightarrow 2} =  \frac{1}{2!}\int d[PS]_2\frac{d\sigma^{2 \rightarrow 2}}{d[PS]_2}\tilde{S}_2(p_1,p_2)
\end{equation}
and
\begin{equation}
\sigma\langle E_T \rangle_{p_0,\Delta Y}^{2\rightarrow 3} = \frac{1}{3!}\int d[PS]_3\frac{d\sigma^{2 \rightarrow 3}}{d[PS]_3}\tilde{S}_3(p_1,p_2,p_3),
\end{equation}
we have integrated the delta functions away, so that our final measurement functions can be written as
\begin{equation}
\label{eq:s2final}
\tilde S_2(p_1,p_2) = \biggl [\epsilon(y_1) + \epsilon(y_2)\biggr ]p_{T2}\theta(p_{T2} \geq p_0)
\end{equation}
and 
\begin{equation}
\label{eq:s3final}
\begin{split}
\tilde S_3(p_1,p_2,p_3) &= (\epsilon(y_1)p_{T1} + \epsilon(y_2)p_{T2} + \epsilon(y_3)p_{T3})\\
&\theta(p_{T1}+p_{T2}+p_{T3} \geq 2p_0)\\
&\theta(\epsilon(y_1)p_{T1} + \epsilon(y_2)p_{T2} + \epsilon(y_3)p_{T3} \geq \beta p_0). 
\end{split}
\end{equation}
Naturally, also the measurement functions $\tilde{S_2}$ and $\tilde{S}_3$ fulfil the IR/CL-safety criteria and thus the $E_T$ moment $\sigma\langle E_T \rangle_{p_0,\Delta Y}$ is a well-defined IR/CL safe quantity to compute.

As a straightforward improvement of the EKRT framework, we now employ the NLO EPS09 nPDFs \cite{EPS09} and CTEQ6M \cite{CTEQ6M:2002} PDFs. 
In fixing the renormalization scale $\mu_R$ and factorization scale $\mu_F$, we follow the common practise and choose them to be equal, $\mu_R = \mu_F = \mu$. We set $\mu$ to be proportional to the hardness of the collision, i.e. to 
the total transverse momentum produced in the hard process, regardless of the partons being in $\Delta Y$ or not:
\begin{equation}
\begin{split}
2 \rightarrow 2:~~& \mu = N_{\mu}(p_{T1}+p_{T2})/2 = N_{\mu}p_T \\
2 \rightarrow 3:~~& \mu = N_{\mu}(p_{T1}+p_{T2}+p_{T3})/2 ,
\end{split}
\end{equation}
where we choose $N_{\mu}=1$.  Note that this choice of $\mu$ is IR/CL safe, as is required for the exact cancellation of the divergences. 

Thus, for the very first time, we are now able to compute the minijet $E_T$ production in $A$+$A$ collisions as rigorously as currently possible to NLO pQCD, and supplement this calculation with the saturation of $E_T$ as described above. We will also study how the nPDF uncertainties, described by the error sets in EPS09, propagate into the computed $E_T$, and thus to the hydrodynamical initial conditions. 

\subsection{Hydrodynamical setup}

In this work we use ideal hydrodynamics to evolve the QCD-matter initial state given by the
EKRT model to final state particles. We solve the equations
\begin{equation}
  \partial_\mu T^{\mu\nu} = 0,
\end{equation}
where $T^{\mu\nu} = (\epsilon + P) u^\mu u^\nu - P g^{\mu\nu}$ is the energy-momentum
tensor, $\epsilon$ is energy density, $P$ is pressure and $u^\mu$ is the fluid
four-velocity. In this work we are only interested in the multiplicity and $p_T$-spectra
in central $A$+$A$ collisions and thus we can make some approximations in order to speed up
the calculations. Firstly, we assume longitudinal boost-invariance
which is a good approximation when discussing results for
mid-rapidities. Secondly, since we limit our studies to central collisions, we
can assume azimuthal symmetry and thus the original 3+1-dimensional numerical problem
is reduced to 1+1 dimensions. In addition, we have set the baryochemical potential
to zero. We employ the SHASTA algorithm \cite{Boris} for solving the hydrodynamical equations.
We use s95p-PCE-v1 equation of state (EoS) \cite{Huovinen:2009yb} (PCE for partial chemical equilibrium)
to close the set of equations. This EoS has a chemical freeze-out at
$T_{\rm chem} = 150$~MeV.

During the hydrodynamical evolution we construct an isothermal freeze-out hypersurface
$\Sigma_\mu$ using a criterion $T_f = 120$~MeV, which was chosen so that we get a reasonable
agreement with the measured $p_T$ spectra at RHIC. We assume that the kinetic freeze-out happens
instantaneously on this surface and the fluid is converted to particles using the Cooper-Frye
formula \cite{Cooper:1974mv}. After we have obtained the thermal particle spectra, we sample
particle ensembles from them in the same way as was done in Ref.~\cite{Holopainen:2010gz}.
The rapidity of the particle is taken from a flat rapidity distribution using an interval
$|y|<3$. Strong and electromagnetic 2- and 3-particle decays are then done one particle
at a time.

Since the transversally-averaged EKRT model considered here does not fix the transverse profile for
the produced initial energy density, we have employed different profiles to get a handle on the
uncertainties related to the profile. Our choices are binary collision and wounded
nucleon (WN) profiles from the optical Glauber model. With the BC profile, the
initial state is obtained as in Eq.~\eqref{eq:enerden} and with the WN profile
the initial energy density is
\begin{equation}
\begin{split}
  \epsilon(s) = &K_{\rm WN} \Big( T_A(s)(1-\exp[-\sigma_{NN}T_B(s)]) \\
  &+ T_B(s)(1-\exp[-\sigma_{NN}T_A(s)]) \Big),
\end{split}
\label{eq:WNprofile}
\end{equation}
where the overall normalization constant $K_{\rm WN}$ is fixed so that we have the same
amount of energy, $E_T(p_{\rm sat})$,  as with the BC profile. Since the entropy is obtained by converting
the energy density to entropy density using the EoS, the total amount of entropy, and thus
also the final multiplicity, is slightly different with the WN profile than with the BC profile.

The hydrodynamic initial time is always taken to be $\tau_0 = 1/p_{\rm sat}$. This relation contains a further ${\cal O}(1)$ proportionality constant which we set to unity throughout this work. For the phenomenology, the value of this constant is obviously correlated with the parameters $K_{\rm sat}$ and $\beta$ but a quantitative investigation of this is left for future studies.

\section{Results}
\label{sec:results}

Next we discuss the results from the present NLO-improved  pQCD + saturation + hydrodynamics framework. We start with the computed NLO minijet $E_T$, charting in particular the effects of the free parameters $\beta$ and $K_{\rm sat}$, as well as the NLO nPDF uncertainties. We demonstrate how the $E_T$-saturation works at RHIC and LHC, and what is the effect of the above uncertainties on saturation, i.e. on the hydrodynamic initial conditions. After understanding these, we fix a specific ($\beta$, $K_{\rm sat}$) combination, and compare our results with the LHC and RHIC data on charged-particle multiplicities as well as on identified-hadron $p_T$ spectra. The propagation of the nPDF errors 
into our multiplicity results is charted in detail. 
As will be seen, the outcome for the predictive power of the improved EKRT approach is quite encouraging.

\begin{figure}[hbt]
\hspace{-1.5cm}
\vspace{-1.5cm}
\begin{center}
\epsfig{file=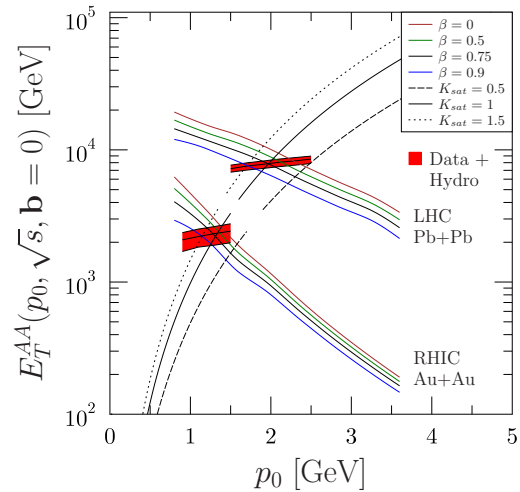, width=9.5cm}
\end{center}
\caption{\label{ETSATHYDRO}(color online) The NLO minijet transverse energy  $E^{AA}_{T}$ produced in 0-5\% central $A$+$A$ collisions in the mid-rapidity region $\Delta Y = 1$, as a function of the $p_T$ cut-off $p_0$. Upper set of the $E_T$ curves:  LHC Pb+Pb at $\sqrt{s_{NN}} = 2.76$ TeV with $\beta=$ 0, 0.5, 0.75 and 0.9. Lower set: RHIC Au+Au at $\sqrt{s_{NN}} = 200$ GeV with the same values of $\beta$.
The rising curves  are $K_{\rm sat} R_{A_{\rm eff}}^2 p_0^3 \Delta Y$, the r.h.s. of Eq.~\eqref{eq:newEKRTsat}, with $K_{\rm sat}=$ 0.5, 1 and 1.5, and $A_{\rm eff}=193\,(181)$ for the LHC (RHIC). The red bands labelled as "Data+Hydro" show how the measured multiplicities translate into the initial states according to ideal hydrodynamics. 
}
\end{figure}

Figure \ref{ETSATHYDRO} shows the average NLO minijet transverse energy  $E^{AA}_{T}$ produced in 0-5\% central $A$+$A$ collisions in the rapidity acceptance $\Delta Y = 1$ as a function of the cut-off $p_0$. We consider Au+Au collisions at the RHIC energy $\sqrt{s_{NN}} = 200$ GeV  and Pb+Pb collisions at the LHC energy $\sqrt{s_{NN}} = 2.76$ TeV. The centrality selection is simulated by considering a central $A_{\rm eff}$+$A_{\rm eff}$ collision of an effective nucleus, $A_{\rm eff} = 181$ at RHIC and $A_{\rm eff} = 193$ at the LHC, as explained in detail in Ref.~\cite{ERRT}. The NLO $E^{AA}_{T}$ curves are computed from Eqs. \eqref{eq:ET} and \eqref{eq:sigmaET} using the measurement functions \eqref{eq:s2final} and \eqref{eq:s3final} with the parameter $\beta =  0$, 0.5,  0.75 and 0.9. As seen in the figure, an increasing $\beta$, which cuts the accepted phase space via the restriction $E_T\ge \beta p_0$, can reduce the amount of the produced $E_T$ as much as by a factor $\sim 2$.

The rising curves  in Fig.~\ref{ETSATHYDRO} are the right-hand side of the saturation equation \eqref{eq:newEKRTsat}, $K_{\rm sat} R_{A_{\rm eff}}^2 p_0^3 \Delta Y$, with the proportionality constant $K_{\rm sat} =$ 0.5, 1, and 1.5.\footnote{The $A_{\rm eff}$ is different for RHIC and LHC, hence the splitting of these curves in the figure.}
Saturation thus takes place when these curves cross those for $E_T$. As the figure demonstrates, the values of $\beta$ and $K_{\rm sat}$ are correlated: the same $E_T$ can be obtained with many different $(\beta,K_{\rm sat})$ pairs. External input -- experimental data -- is needed to resolve the best allowed values for these parameters. To get a hold on this, the red bands in Fig.~\ref{ETSATHYDRO} are to indicate which $E_T^{AA}(p_0)$
would be needed according to our hydrodynamic prescription, assuming $\tau_0=1/p_0$ and the BC energy-density profile, to reproduce the experimentally measured multiplicities at RHIC and LHC.\footnote{Thus, the input for the red bands contains only hydrodynamics and experimental data, no pQCD or saturation.} The solid lines inside the red bands are computed from the  statistically weighted averages of the measured charged-particle multiplicities at the LHC (ALICE \cite{Aamodt:2010pb} and CMS \cite{CMSMUL:2011}) and RHIC (PHENIX \cite{PHENIXMUL:2005}, STAR \cite{STARMUL:2009} and BRAHMS \cite{BRAHMSMUL:2002}). The red error bands 
correspond to the largest (smallest) values of the experimental upper (lower) limits of multiplicities, mapped now into the QCD-matter initial state through ideal hydrodynamics. 

We can also see from the figure that if we choose $\beta =0$ and tune $K_{\rm sat}$ to fit the RHIC data, we overshoot the "experimental" error band at the LHC -- this was a problem with the old EKRT setup, where the $\sqrt s$ scaling of the multiplicity became too strong. Interestingly, however, we observe that one effect of our hardness-parameter $\beta$ is the taming of this $\sqrt s$ scaling: if we choose, as perhaps the most natural choice would be, $K_{\rm sat}=1$ (the solid rising curve) and $\beta=0.75$ (the solid black $E_T$ curves), we match the average LHC multiplicity perfectly and agree very nicely also with the RHIC average multiplicity. As seen in the figure, the acceptable values of $\beta$ and $K_{\rm sat}$ are clearly correlated: with a larger $\beta$ a larger value of $K_{\rm sat}$ is needed.

\begin{figure}[hbt]
\hspace{-1.5cm}
\vspace{-1.5cm}
\begin{center}
\epsfig{file=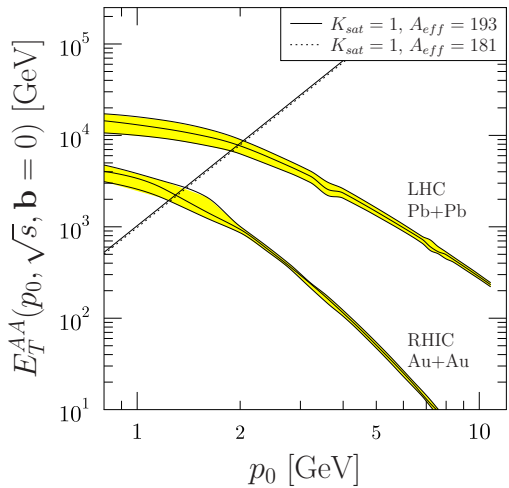, width=9.5cm}
\end{center}
\caption{\label{ET_EPS09ERR_075}(color online) The computed NLO minijet $E^{AA}_T$ in $\Delta Y=1$ with fixed $\beta = 0.75$,  for RHIC Au+Au collisions at $\sqrt{s_{NN}} = 200$ GeV and LHC Pb+Pb collisions at $\sqrt{s_{NN}} = 2.76$ TeV, including the nPDF uncertainties from the 30 EPS09 error sets. The  $A_{\rm eff}$s simulate the 5\% most central collisions.}
\end{figure}

Figure \ref{ET_EPS09ERR_075} shows the computed NLO $E^{AA}_T$ in $\Delta Y =1$ for fixed $\beta = 0.75$ at RHIC (lower bands) and LHC (upper bands). The straight lines in this log-log plot are again the r.h.s. of Eq.~{\eqref{eq:newEKRTsat} with a fixed $K_{\rm sat}=1$}, and the $A_{\rm eff}$s simulating the 0-5\% centralities at RHIC and LHC. 
The black lines inside the error bands are the $E^{AA}_T$
computed by using the EPS09 best fit $S^0$.
The yellow error bands show the uncertainty which originates from the nPDF uncertainties, computed using the 30 error sets in EPS09 according to the prescription given in \cite{EPS09}: 
\begin{equation}
\label{eq:epseterr}
\begin{split}
(\Delta E^{AA}_T)^{\pm} = \sqrt{\sum_{k}\left (\max (E^{AA}_T)^{\pm}_{k} \right )^2},
\end{split}
\end{equation}
where
\begin{equation}
\begin{split}
\max(E^{AA}_T)^{+}_{k} = \max \{ &E^{AA}_T(S^{+}_{k})-E^{AA}_T(S^0),\\
&E^{AA}_T(S^{-}_{k})-E^{AA}_T(S^0),0 \} 
\end{split}
\end{equation}
and
\begin{equation}
\begin{split}
\max(E^{AA}_T)^{-}_{k} = \max \{ &E^{AA}_T(S^0)-E^{AA}_T(S^{+}_{k}),\\
&E^{AA}_T(S^0)-E^{AA}_T(S^{-}_{k}),0 \},
\end{split}
\end{equation}
where $E^{AA}_T(S^{\pm}_{k})$ denotes the value of the $E^{AA}_T$ computed with the set $S^{\pm}_{k}$ for $k=1,\ldots,15$.

In Fig.~\ref{ET_EPS09ERR_075}, we observe that at the LHC the error bands shrink consistently towards higher $p_0$. This is due to the DGLAP scale evolution of nuclear gluon PDFs, which rapidly decreases the shadowing of small-$x$ gluons as well as their uncertainties. Also note here that the minimum scale of both the CTEQ6M PDFs and EPS09 nuclear effects is $Q_0=1.3$ GeV, and that for the nPDFs below $Q_0$ we have used those at $Q_0$ for simplicity. Thus the results at $p_0<1.3$ GeV in Figs.~\ref{ETSATHYDRO} and \ref{ET_EPS09ERR_075} are not fully consistent with the DGLAP evolution. However, with the choice $\beta = 0.75$ and $K_{\rm sat}=1$,  we conveniently have $p_{\rm sat}>Q_0$ at saturation also at RHIC (see Table~1 ahead).

For the RHIC results in Fig.~\ref{ET_EPS09ERR_075}, we notice a more non-monotonic behaviour in the widths of the corresponding error bands, as well as a few-percent numerical uncertainty at $1.3\lesssim p_0\lesssim 2$~GeV. These can be traced back to the non-trivial behaviour of the large NLO $2\rightarrow2$ contribution and its interplay between the $2\rightarrow3$ contributions which require 6 dimensional MC integrations \cite{ET2}.	Similarly, the small numerical fluctuations seen at the LHC curves originate from these multidimensional MC integrations.  

Furthermore, we can see a general trend that for a fixed $p_0\gtrsim 2$ GeV the error bands are smaller at RHIC than at the LHC. This again is a reflection of the different magnitudes of the uncertainties in the nuclear gluon PDFs at different values of $x$: at RHIC, due to the smaller $\sqrt{s_{NN}}$, one probes larger values of $x$ where the shadowing uncertainties are smaller than in the smaller-$x$ region probed at the LHC.

Finally, and perhaps most importantly, Fig.~\ref{ET_EPS09ERR_075} shows that the rigorous NLO $E_T$ computation itself seems to be well under control even at the rather low perturbative scales at saturation, and, that the nPDF-originating uncertainties remain rather modest, ca. 15 \% at saturation both at the LHC and RHIC (see Table~1). 
 
\begin{figure}[htb]
\begin{center}
\epsfig{file=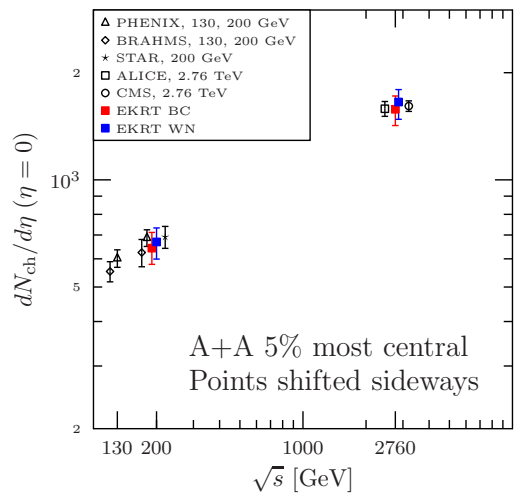, width=7.5cm}
\end{center}
\caption{\label{dNdeta_vs_s_errsets}(color online) Charged-particle multiplicity $dN_{\rm ch}/d\eta$ at $|\eta| < 0.5$ in 0-5\% most central $\sqrt{s_{NN}} = 2.76$ TeV Pb+Pb collisions at the LHC and $\sqrt{s_{NN}} = 200$ GeV Au+Au collisions at RHIC, obtained from NLO pQCD + saturation + hydrodynamics using the BC and WN initial energy-density profiles (filled red and blue squares, correspondingly), and a comparison with the ALICE and CMS data (LHC) Refs. \cite{Aamodt:2010pb,CMSMUL:2011} and PHENIX, STAR and BRAHMS data (RHIC) Refs. \cite{PHENIXMUL:2005,STARMUL:2009,BRAHMSMUL:2002}. Notice that most of the points have been shifted sideways to improve their visibility.
}
\end{figure}
In Fig.~\ref{dNdeta_vs_s_errsets} we show the computed
charged-particle multiplicity $dN_{\rm ch}/d\eta$ at $|\eta| < 0.5$ in 0-5\% most central $\sqrt{s_{NN}} = 2.76$ TeV Pb+Pb collisions at the LHC and $\sqrt{s_{NN}} = 200$ GeV Au+Au collisions at RHIC, obtained in the current framework of  NLO pQCD + saturation + hydrodynamics with $\beta = 0.75$ and $K_{\rm sat} = 1$, using the BC  and WN initial energy-density profiles.
Also the comparison with the ALICE \cite{Aamodt:2010pb} and CMS \cite{CMSMUL:2011} data (LHC) and PHENIX \cite{PHENIXMUL:2005}, STAR \cite{STARMUL:2009} and BRAHMS \cite{BRAHMSMUL:2002} data (RHIC) is shown. 

The error bars to our results in Fig.~\ref{dNdeta_vs_s_errsets} have been obtained as follows:
First, we calculate the NLO  $E^{AA}_T$ in $\Delta Y=1$ for RHIC and the LHC by fixing $\beta = 0.75$ and using the best fit and the 30 error sets of EPS09. Second, we solve the saturation equation \eqref{eq:newEKRTsat} and determine $p_{\rm sat}$ with a fixed $K_{\rm sat} = 1$ for all the 31 different NLO $E^{AA}_T$ results, and thus at saturation we find 31 different $(E^{AA}_T(p_{\rm sat}),p_{\rm sat})$ pairs. Third, we construct the hydrodynamical initial conditions, 
$\epsilon(s,\tau_0,\sigma\langle E_T \rangle_{p_{\rm sat}})$  at $\tau_0 = 1/p_{\rm sat}$ (see Eqs.~\eqref{eq:enerden} and \eqref{eq:WNprofile})
and calculate $dN_{\rm ch}/d\eta$	for every EPS09 set separately. Finally, using these numbers, we compute the theoretical error bars 
using the EPS09 prescription of Eq.~\eqref{eq:epseterr}. We can see that the error bars to our results
are slightly larger than the experimental error bars at the LHC, and of the same order as those at RHIC, indicating again that the calculation presented here is not suffering from a large gluon nPDF uncertainty.

Thus, from Fig. \ref{dNdeta_vs_s_errsets} we conclude that our NLO pQCD + saturation + (ideal)hydrodynamics framework 
reproduces the measured RHIC and LHC multiplicities quite nicely and the $\sqrt{s_{NN}}$ scaling seems to work very well. The transverse-profile uncertainty in the computed multiplicity is a few percent and the nPDF-related uncertainty about $\pm$10\%, for fixed $\beta$ and $K_{\rm sat}$,
 both at RHIC and LHC. Most importantly, since it is possible to keep the values of $\beta$ and $K_{\rm sat}$ fixed, we can conclude that our framework has some definite predictive power.

\begin{table}[ht]
\centering
\begin{tabular}{c c c}
\hline\hline
  & RHIC Au+Au & LHC Pb+Pb  \\ [0.5ex]
\hline
$\sqrt{s_{NN}}$ [GeV] & 200 & 2760\\ [1ex]
$A_{eff} $       & 181 & 193 \\ [1ex]
$T_{A_{\text{eff}}A_{\text{eff}}}({\bf 0})$ [1/mb] & 26.0 & 28.5 \\ [1ex]
$p_{\text{sat}}$ [GeV] & $1.31\,^{0.07}_{0.06}$ & $1.96\,^{0.08}_{0.10}$ \\ [1ex]
$E^{A_{\text{eff}}A_{\text{eff}}}_{T}(p_{\text{sat}})$  [GeV]   & $2202\,^{369}_{272}$  & $7720\,^{973}_{1110}$       \\ [1ex]
$\sigma\langle E_T \rangle(p_{\text{sat}})$  [mbGeV]   & $84.6\,^{14.2}_{10.4}$  & $271\,^{34}_{39}$       \\ [1ex]
$\tau_0$ [fm]    & $0.151\,^{0.006}_{0.008}$ & $0.100\,^{0.005}_{0.004}$ \\  [1ex]  
$dN_{\rm ch}/ d\eta$ (BC)  & $643\,^{69}_{64}$  & $1579\,^{142}_{158}$     \\ [1ex]
$dN_{\rm ch}/ d\eta$ (WN)  & $669\,^{64}_{70}$  & $1652\,^{139}_{175}$     \\ [1ex]
\hline 
\end{tabular}
\caption{The collision parameters, the obtained NLO pQCD + saturation key results 
for fixed $\beta =0.75$ and $K_{\text{sat}} = 1$, the corresponding hydrodynamical input and final results for the charged-particle multiplicities.
}
\label{table:nonlin}
\end{table}

Table~1 collects the collision parameters as well as our results for $p_{\rm sat}$, $E_T^{AA}(p_{\rm sat})$ at saturation, the corresponding hydrodynamic inputs $\tau_0$ and $\sigma\langle E_T \rangle$, as well as the multiplicities. The errors 
shown for these quantities (upper and lower limits separately) have been computed based on the 30 error sets of EPS09.

\begin{figure}[ht]
\centering
\epsfig{file=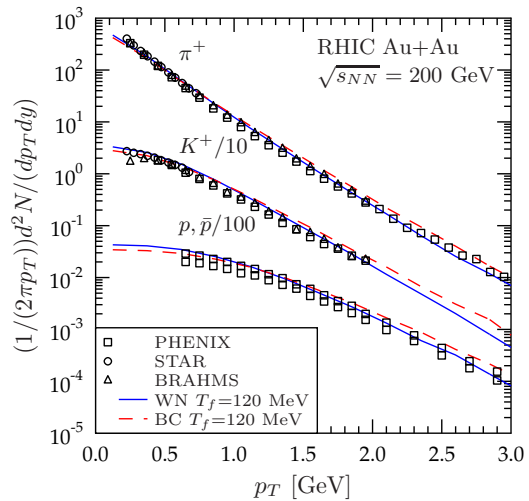, width=7.5cm}
\caption{\label{spectra_rhic}(color online) 
The $p_T$ spectra of $\pi^+$, $K^+$, $p$ and $\bar p$ in 0-5 \% most central Au+Au collisions at
$\sqrt{s_{NN}} = 200$ GeV at RHIC, computed from NLO pQCD + saturation + hydrodynamics with the WN profile
(solid blue lines) and BC profile (dashed red lines), and measured by the PHENIX \cite{PHENIXSPECT:2004} (squares), STAR \cite{STARSPECT:2009} (circles) and BRAHMS (triangles) \cite{BRAHMSSPECT:2004} experiments.
The calculated spectra and the PHENIX data are without the hyperon feed-down contributions.}
\end{figure}

Figure \ref{spectra_rhic} shows the $p_T$ spectra of $\pi^+$, $K^+$, $p$ and $\bar p$ in 0-5 \% most central Au+Au collisions at $\sqrt{s_{NN}} = 200$ GeV at RHIC, computed from NLO pQCD + saturation + hydrodynamics with 
$\beta=0.75$ and $K_{\rm sat}=1$, and the WN and BC initial energy-density profiles. These results thus correspond to the computed RHIC charged-particle multiplicities shown in Fig.~\ref{dNdeta_vs_s_errsets}, and we see that with the adopted PCE EoS and $T_f=120$ MeV, both the particle multiplicities and their $p_T$ spectra measured by PHENIX \cite{PHENIXSPECT:2004}, STAR \cite{STARSPECT:2009}
and BRAHMS \cite{BRAHMSSPECT:2004} become reproduced quite well. Note, however, that since we do not include any net-baryon number in the hydrodynamical calculation, we cannot address the difference between the protons and antiprotons here.

\begin{figure}[ht]
\centering
\epsfig{file=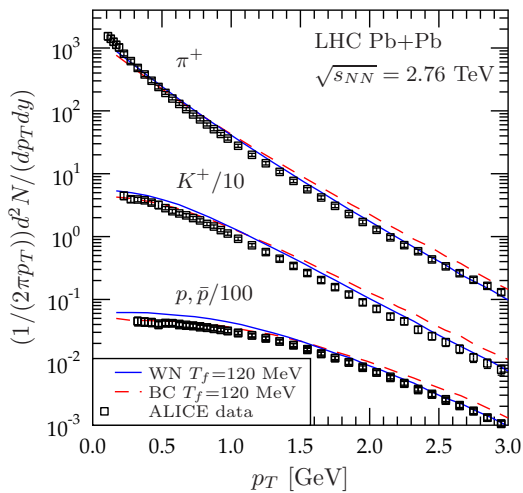, width=7.5cm}
\caption{\label{spectra_lhc}(color online) 
The $p_T$ spectra of $\pi^+$, $K^+$, $p$ and $\bar p$ in 0-5 \% most central Pb+Pb collisions at
$\sqrt{s_{NN}} = 2.76$ TeV at the LHC, computed from NLO pQCD + saturation + hydrodynamics with the WN profile
(solid blue lines) and BC profile (dashed red lines), and measured by the ALICE \cite{ALICESPECT:2012} (squares) experiment.}
\end{figure}

Figure \ref{spectra_lhc} shows the corresponding identified particle $p_T$ spectra for 0-5 \% most central Pb+Pb collisions at $\sqrt{s_{NN}} = 2.76$ TeV, computed in our improved NLO EKRT framework with the same $T_f=120$~MeV as at
RHIC, compared with the very recently published spectra from the ALICE experiment \cite{ALICESPECT:2012}. Again, the computed spectra correspond to the computed LHC charged-particle multiplicities shown in Fig.~\ref{dNdeta_vs_s_errsets}. Very interestingly, we find that also at the LHC  we can reproduce both the particle multiplicities and their $p_T$ spectra quite well. Especially, we would like to emphasize that the computed results are not a fit to these ALICE data but a prediction in the following way: First, we fixed the parameters $\beta=0.75$ and $K_{\rm sat}=1.0$ by requiring that the measured charged-particle multiplicity is reproduced at the LHC (Fig.~\ref{ETSATHYDRO}, the red band). Simultaneously, also the RHIC multiplicity comes out well.
Then we fixed the kinetic decoupling temperature $T_f=120$~MeV on the basis of Fig.~\ref{spectra_rhic}, and kept it fixed when computing the LHC spectra in Fig.~\ref{spectra_lhc}.\footnote{The charged-particle multiplicities are not sensitive to $T_f$.} Regarding the initial energy-density profile uncertainty, we can conclude based on Figs.~\ref{spectra_rhic} and \ref{spectra_lhc} that a flatter WN profile would seem to reproduce the identified particle $p_T$ spectra better than the steeper BC profile.

\newpage
\section{Discussion and outlook}
\label{sec:outlook}

In this work, we introduced an improved formulation of the pQCD + saturation (EKRT) model \cite{EKRT}, where the saturation of the gluon minijets occurs for the transverse energy, $E_T$, rather than for the number of gluons. Since the minijet $E_T$ is a quantity safe from the infrared and collinear singularities, it can be computed rigorously to NLO pQCD as a function of the minijet transverse-momentum cut-off $p_0$. The IR/CL safe minijet $E_T$ in a chosen acceptance region is defined by measurement functions analogous to jet physics. We showed that in fact a set of infinitely many equally possible IR/CL-safe measurement functions for the $E_T$ can be formulated in terms of a hardness parameter $\beta$. In computing the minijet $E_T$ to NLO with these new measurement functions,  we also updated the framework of Refs.~\cite{ET1,ET2} with the EPS09 NLO nPDFs \cite{EPS09}, and studied the propagation of the nPDF uncertainties to the minijet $E_T$ with the 30 error sets of EPS09. Thus, in the present study the computation of the minijet $E_T$ production is performed for the very first time genuinely and consistently to NLO in $A$+$A$ collisions.

In the more phenomenological part of the study, we formulated a dynamical saturation criterion for the minijet $E_T$, which solves the open issues of the old EKRT-model regarding the appearance of the rapidity acceptance interval $\Delta Y$ and powers of $\alpha_s$  in the saturation criterion. The outcome is the EKRT-like geometrical saturation criterion for the collinearly factorized minijet $E_T$ production, containing one unknown proportionality constant $K_{\rm sat}$.
Applying this saturation criterion to our NLO computation for the average minijet $E_T$ production in 5 \% most central $A$+$A$ collisions at RHIC and LHC, we determined the saturation scales $p_{\rm sat}$ and the amounts of $E_T$ produced into a mid-rapidity unit at saturation. In particular, we demonstrated how these depend on the model parameters $\beta$ and $K_{\rm sat}$, and we also quantifed the EPS09-originating error bars to these quantities.  

Converting the saturated minijet $E_T$ into the QCD matter energy density at $\tau_0=1/p_{\rm sat}$, thus assuming thermalization at formation, and adopting either a BC or a WN transverse profile, we obtained initial conditions for the hydrodynamic evolution. In this baseline study, where our main goal is to chart the general features and predictive power of the NLO-improved pQCD+ saturation + hydrodynamics modeling rather than a detailed fitting of the data, we restricted ourselves to boost-invariant ideal hydrodynamics with a state-of-the art PCE EoS \cite{Huovinen:2009yb}. What we believe is particularly useful for understanding the initial-state phenomenology here, is that we concretely showed how the measured charged-particle multiplicities translate into the QCD-matter initial conditions through hydrodynamics (the red bands in Fig.~\ref{ETSATHYDRO}). Using this mapping, we could directly see which parameter pairs $(\beta, K_{\rm sat})$ -- if any - would reproduce the measured LHC and RHIC charged-particle multiplicities simultaneously. Remarkably, the outcome is that (at least in this ideal-hydrodynamics framework) such a simultaneous reproduction of multiplicities is very well possible, and that we can determine the acceptable range of values for these correlated parameters.
 Using then one possible parameter combination, $\beta=0.75$ and $K_{\rm sat}=1$, we computed the identified particle $p_T$ spectra for the 5 \% most central $A$+$A$ collisions at RHIC and LHC, fixing the kinetic decoupling temperature from the measured RHIC spectra. Again, the outcome is that the particle multiplicities and $p_T$ spectra are very nicely reproduced simultaneously both at RHIC and the LHC --- emphasizing the fact that the  identified hadron $p_T$ spectra for the LHC Pb+Pb collisions which we obtained here, is a prediction, and not a fit to the ALICE $p_T$ spectra. 

To summarize, we have shown that the NLO-improved pQCD + saturation + ideal-hydrodynamics is a viable model for describing particle production in central heavy-ion collisions at the LHC and RHIC. We have quantified the key parameters of the approach and studied the propagation of different uncertainties into the hadron multiplicities and $p_T$ spectra. Most importantly, our results indicate that the framework has definite predictive power: The key-parameters of the NLO pQCD calculation ($\beta$) and saturation ($K_{\rm sat}$) can be fixed -- not uniquely but in a correlated manner -- based on the measured charged-particle multiplicity in $A$+$A$ collisions at \textit{one} given cms-energy $\sqrt{s_{NN}}$,  and the kinetic and chemical freeze-out temperatures ($T_f,T_{\rm chem}$)  based on the measured identified particle $p_T$ spectra and multiplicities at the \textit{same} cms-energy. After this, predictions for other $\sqrt{s_{NN}}$ and $A$ are can be computed. 

The results presented in this paper are quite encouraging for the rather obvious further developments of the framework. Following Ref.~\cite{EKT1}, we should next extend the study to non-central collisions. This requires a localization of the saturation criterion, which in turn calls for the spatial dependence of the nPDFs in the colliding nuclei. Thanks to the recent developments, these tools now exist \cite{EPS09s}, so that also the localized saturation study can now be brought consistently to NLO. A localized saturation will also fix the transverse profiles of the initial energy densities in the saturated interior of the system, thus considerably decreasing the profile uncertainty.  Ultimately, to compute initial conditions for the event-by-event hydrodynamics, such a localized study should be performed event-by event, using a Monte Carlo simulation which accounts for the fluctuations both in the number of binary $NN$ collisions and in the minijet multiplicity ($E_T$) from one $NN$ collision to another. 

Another very interesting line of further studies is the inclusion of dissipation, viscous corrections, to our hydrodynamic description. Dissipation will somewhat increase the entropy during the hydrodynamic evolution of the system, and the viscous corrections also affect the hadron $p_T$ spectra. Promisingly, Fig.~\ref{ETSATHYDRO} indicates that that there indeed is room for such an entropy increase, with suitably chosen values of
the parameters $\beta$ and $K_{\rm sat}$. 
It will be very interesting to see whether the present framework still retains its predictive power also in the presence of dissipation. 

\acknowledgments
This work was financially supported by the Wihuri foundation (R.P.), the national Graduate School of
Particle and Nuclear Physics, the Magnus Ehrnrooth foundation (K.J.E.), the Academy of Finland, K.J.E.'s project 133005 and T. Lappi's project 141555.  H.H. was supported by the Extreme Matter Institute (EMMI).
We thank T.~Lappi, T.~Renk, H.~Niemi, K.~Kajantie, H.~M\"antysaari and B.~M\"uller for useful discussions, and CSC-IT Center Science for supercomputing time.

\end{document}